\documentclass[11pt]{article}

\usepackage[style=authoryear]{biblatex}
\usepackage{hyperref}
\hypersetup{colorlinks,allcolors=black}
\usepackage{xcolor}
\usepackage{soul}
\usepackage{pifont}

\usepackage{graphicx} % Required for inserting images
\usepackage[T1]{fontenc}
\usepackage{setspace}
\usepackage[letterpaper, portrait, margin=1in]{geometry} % Set page size and margin

\title{{\textbf{Functionality Locality, Mixture \&\\ Control = Logic = Memory}}}
\author{Xiangjun Peng\\
Succincter\\
\href{mailto:shiangjun@succincter.com}{shiangjun@succincter.com}\\
}

\date{}

\begin{document}

\maketitle

%%%%%% -- PAPER CONTENT STARTS-- %%%%%%%%

\begin{abstract}
  \noindent
  This work provides new insights and constructs to the field of computer architecture and systems, and these insights are expected to be useful for the broad software stack. First, this work introduces Functionality Locality: this form of Functionality Locality shows that functionalities can be changed with a single piece of information, by solely changing the access order. This broadens the scope of ``principle of locality", which originally includes spatial and temporal locality. Second, this work coins the term Mixture, by incorporating the layout-directed functionalities with the original quantifiers such as scalar and vector. The implications of Mixture significantly expands new understanding of quantifiers, and this work identifies several important ones (from the author perspective). Third, with Functionality and Mixture, this work identifies the principle ``Control = Logic = Memory", and provides a revisit to Von Neumann architectures and Harvard architectures. This centers the focus on the memory, and brings further guidelines on memory-centric architectures with a new analytic framework. Fourth, this work discusses several important implications from this work in a variety of aspects. 
  
\end{abstract}

\newpage
\tableofcontents
\newpage

\begin{singlespace}

\section{Functionality Locality}

\vspace{-4pt}
\noindent
This section first documents the current status of ``locality of reference" (i.e., the form of ``principle of locality" in this subject) in computer science. Then, this section introduces Functionality Locality, a new form of locality based on the breakthrough of the theory and practice in computer science. Finally, this section delivers the practical example from the newly-delivered form of the locality.

\vspace{-4pt}

\subsection{Spatial Locality and Temporal Locality}

\vspace{-2pt}

\noindent
To the best of the knowledge from the author, the concept of locality shall be originally credited to physics. A quotation can be delivered as follows. ``In physics, the principle of locality states that an object is influenced directly only by its immediate surroundings. A theory that includes the principle of locality is said to be a ``local theory"." These concepts are used in computer science, and rephrased as ``Locality of Reference". This work quotes the following classification, and supplies the evolved types of locality for recent developments.

\vspace{-4pt}
\begin{itemize}
    \item \textbf{Temporal locality.} If at one point a particular memory location is referenced, then it is likely that the same location will be referenced again in the near future. There is temporal proximity between adjacent references to the same memory location. In this case it is common to make efforts to store a copy of the referenced data in faster memory storage, to reduce the latency of subsequent references. Temporal locality is a special case of spatial locality (see below), namely when the prospective location is identical to the present location.
    \vspace{-2pt}

    \item \textbf{Spatial locality.} If a particular memory/storage location is referenced at a particular time, then it is likely that nearby memory locations will be referenced in the near future. In this case it is common to attempt to guess the size and shape of the area around the current reference for which it is worthwhile to prepare faster access for subsequent reference.
\end{itemize}
\vspace{-4pt}

\noindent
With the evolve of theory and practice in computer architecture, there are also following types of locality to be mentioned.

\vspace{-4pt}
\begin{itemize}
    \item \textbf{Branch locality.} If there are only a few possible alternatives for the prospective part of the path in the spatial-temporal coordinate space. This is the case when an instruction loop has a simple structure, or the possible outcome of a small system of conditional branching instructions is restricted to a small set of possibilities. Branch locality is typically not spatial locality since the few possibilities can be located far away from each other.\vspace{-2pt}

    \item \textbf{Equidistant locality} Halfway between spatial locality and branch locality. Consider a loop accessing locations in an equidistant pattern, i.e., the path in the spatial-temporal coordinate space is a dotted line. In this case, a simple linear function can predict which location will be accessed in the near future.
\end{itemize}
\vspace{-4pt}

\noindent
Based on the above classifications, we can conclude that: all concepts can be derived from Spatial Locality. Therefore, the central of our focus lies on Spatial Locality, and it is expected to be consistent with new concepts.

%\vspace{-4pt}

\subsection{Functionality Locality}

%\vspace{-2pt}

\noindent
The key insight of this work is that: all existing understanding on locality only considers how references shall be taken advantage of, without the consideration on the access order of the information. A recent work~(\cite{ArXiv24/Peng}) demonstrates that it is feasible to deliver polymorphic functionalities within a single piece of information. Therefore, it is expected to revisit the local theory (from physics), and how such a revisit plays a role computer architecture. 

%\vspace{-12pt}
\begin{figure}[!h]
    \centering
    \includegraphics[width=0.7\linewidth]{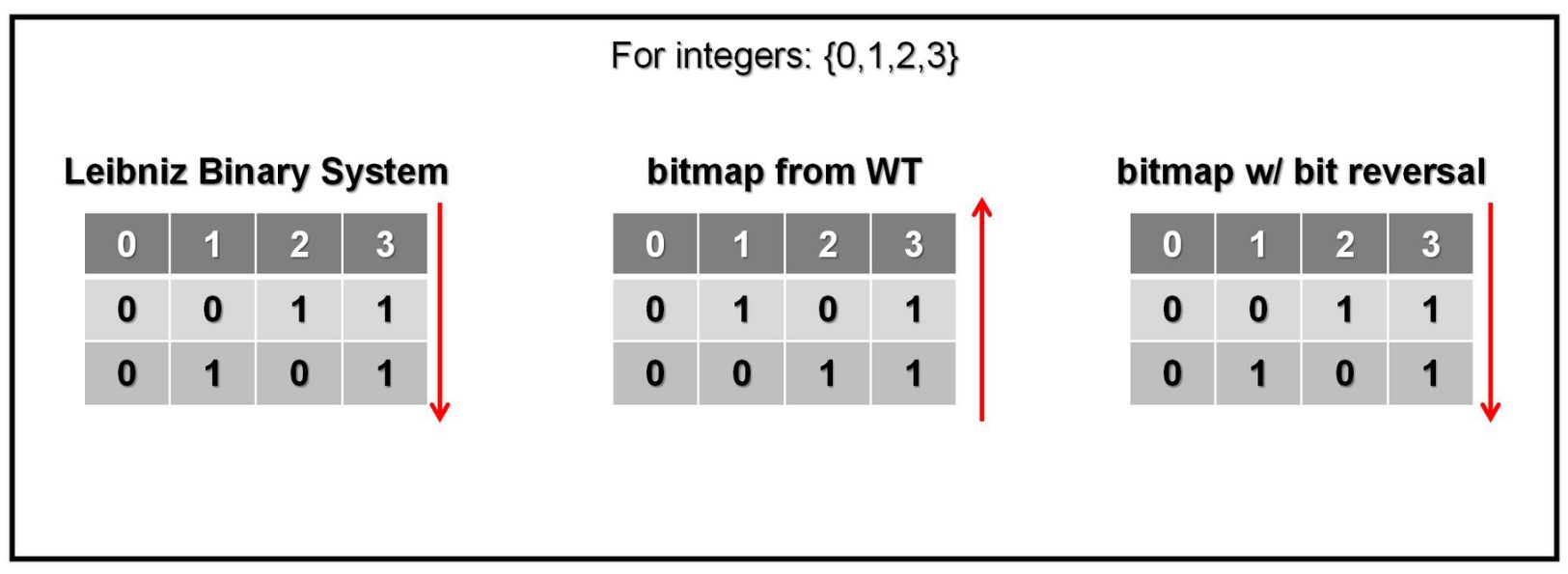}
%    \vspace{-16pt}
    \caption{A pictorial example from~(\cite{ArXiv24/Peng}.}
    \label{fig:revival}
\end{figure}

\noindent
\textbf{Functionality Locality.} It can be defined as: the access order of a single piece of information can determine different functionalities, though the information has no spatial changes. Figure~\ref{fig:revival} demonstrates an example from~(\cite{ArXiv24/Peng}, and it is very straightforward to understand: \ding{202} from top to bottom, the binary system can be used for computation; and \ding{203} from bottom to top, the bitmap is used for queries. Also, it can be understood as: a single piece of information can be used in more-than-one systems, which are originally oriented for different functionalities.

%\vspace{-4pt}

\subsection{Practical Examples of Functionality Locality}
\label{sec:TRA}

%\vspace{-2pt}

\noindent
The discovery of Functionality Locality motivates a retrospection on computer designs, in terms of the bit-parallel (i.e., horizontal) and bit-serial (i.e., vertical) data layouts. To date, modern processors leverage bit-parallel layouts, to maximize the computational power. This substantially causes the issues of data movements between the processor and memory. Therefore, such an issue ignites the trend of Processing Using Memory in DRAM (i.e., the modern memory chips) by~(\cite{ArXiv19/Seshadri-PUM}), and serves as the basis for the recently-discussed paradigm - Memory-Centric Computing~(\cite{ArXiv23/Mutlu-MCC})\footnote{Note that, though the identified idea can be used within the processors, the benefits can not be fully unleashed as explained.}.

%\vspace{-8pt}
\begin{figure}[!h]
    \centering
    \includegraphics[width=0.5\linewidth]{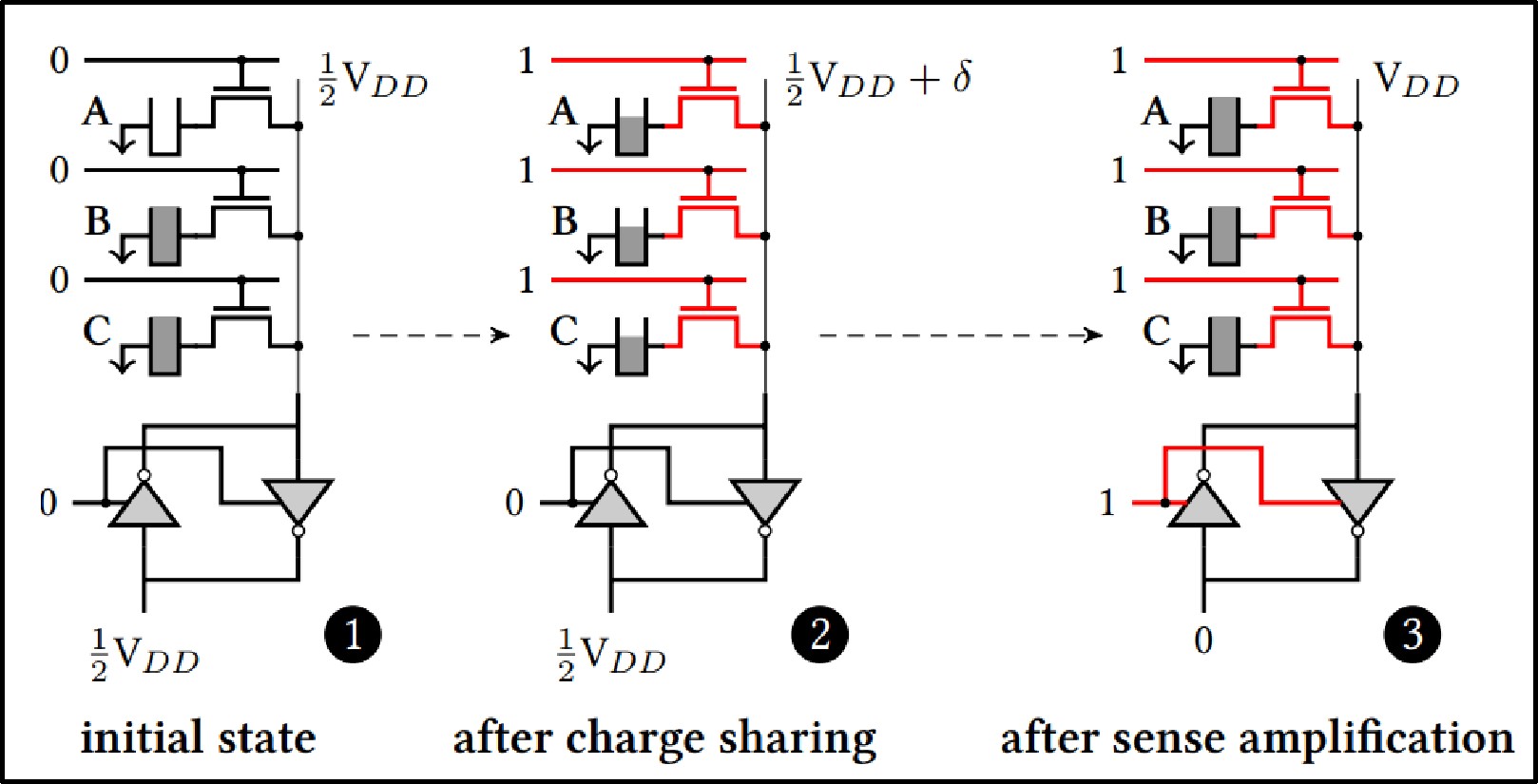}
%    \vspace{-16pt}
    \caption{A pictorial example from~(\cite{ArXiv19/Seshadri-PUM}) on Processing Using DRAM.}
    \label{fig:TRA}
\end{figure}

%\vspace{-12pt}

This work revisits the idea in a short manner, to serve as the foundation for the further discussion. Figure~\ref{fig:TRA} gives out the example: after the three steps, The final state can be written as AB + BC + AC. With some proper changes, the final state is C(A + B) + $\neg$C(AB). Therefore, one can view A and B as the operand bit, and C as the control bit.

\newpage

\section{Mixture}
%\vspace{-2pt}

\noindent
This section captures the implications from Functionality Locality, and coins the term Mixture. This section first describes the definition of Mixture, and describes its relationships with functionalities and bit length. Then, this section depicts the proper settings to combine Mixture with an analytic framework, which consists of (ordered) values, representations (if needed) and information.

%\vspace{-4pt}

\subsection{Definition}

%\vspace{-2pt}

\noindent
Mixture is a bit sequence with the arbitrary length, and its functionalities include compute\footnote{Note that any computations can be formalized using logic operations due to the logic completeness.}, query and move. Therefore, the following guidelines can be derived, based on the length of this sequence, and how the sequence is layouted.

\noindent
\begin{itemize}
    \item When the length is one (namely, only one bit), a Mixture can compute, query and move.
    \item When the length is more than one (namely, there are multiple bits), the functionalities of a Mixture can be weakened due to the sequence layout:
    \begin{itemize}
        \item if horizontally layouted (i.e., bit-parallel), it can only compute and move.
        \item if vertically layouted (i.e., bit-serial), it can only query and move.
    \end{itemize}
\end{itemize}

The above definitions shall cover all existing quantifiers (such as scalar and vector) and bitmaps. However, following the ``Indexes $\approx$ Values" principle~(\cite{ArXiv24/Peng}), a Mixture can compute, query and move without any additional layout transformations.

%\vspace{-4pt}

\subsection{Synergy with an Analytic Framework}

%\vspace{-2pt}

\noindent
\cite{ArXiv24/Peng-VRIC} provides an analytic framework to connect (ordered) values, representations (if needed) and (disordered) information, in a consistently-recursive manner. This substantially motivates a question: can there be the minimal set of them be stored, so that the rest can be fully restored? The definition of Mixture naturally complements with the above requirements: only those ones, which can meet the full functionalities, shall be kept so that the rest can be fully restored. This conjecture can be mathematically explained via Mandelbrot set and/or Julia set, which are recursively used for the formation.

%\vspace{-4pt}

\newpage

\section{Control = Logic = Memory}

%\vspace{-2pt}

\noindent
This section captures the implications from Functionality Locality and Mixture based on the provided example, and provides a revisit two widely-practiced architectures: Von Neumann architecture~(\cite{von2021first}) and Harvard architecture.

%\vspace{-4pt}

\subsection{The ``Control = Logic = Memory" Principle}

\vspace{-2pt}

\noindent
As described in Section~\ref{sec:TRA}, exploiting analog capabilities of three cells delivers a formation as C(A + B) + $\neg$C(AB). This work leverages the perspectives from~(\cite{ArXiv24/Peng-VRIC}), and claims that: the denotations of A, B and C are simply the representations, due to the temporality. In fact, the actual fact is there are interactions between A, B and C at a time slot. Therefore, this work makes a conjecture with this special case, Control can be equalized with Logic, and Logic can be equalized with Memory: namely Control = Logic = Memory.

%\vspace{-4pt}

\subsection{Everything is Memory}

%\vspace{-2pt}

\begin{figure}[!h]
    \centering
    \includegraphics[width=0.7\linewidth]{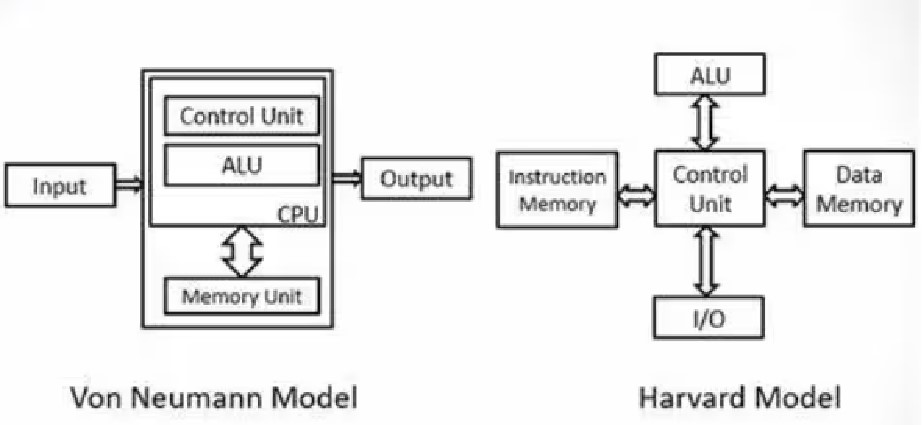}
%    \vspace{-16pt}
    \caption{Von Neumann Model versus Harvard Model.}
    \label{fig:model-compare}
\end{figure}

\noindent
A revisit to existing architectures is delivered in Figure~\ref{fig:model-compare}.

%\vspace{-8pt}

\noindent
With the identified principle, Control = Logic = Memory, it is clear that Harvard model can be understood as a form of specialization of Von Neumann Model, with different levels of the granularity for the memory. Therefore, the conceptual reasoning in this work expands Memory-Centric Computing into Everything is Memory. More importantly, this expansion, along with Mixture (particularly Mandelbrot and/or Julia sets), provides a completely different perspective on self-replication theory~(\cite{von-self-replicate}) (which solely focused on representations). 

\subsection{Memory-Centric Architectures}

\noindent
This work conjectures that, the computer organization shall be used to store all Mixtures, formalized as an analog with the recursively-formalized analytic framework~(\cite{ArXiv24/Peng-VRIC}). Therefore, all components, as memory, are recursively organized and the hierarchy is expected to be asynchronous, in terms of the granularity. This is because: the purposes of recursively-organized memory are to store ((sub)sets of) Mixtures, and they are not necessarily symmetry. It would be an interesting discussion, to see whether the functionalities of such a computer shall be determined in advance. Clearly, it depends on whether the artificial intelligence shall play a god, though nobody can guarantee that the god (if any) truly prefers that human beings shall exist or not.

%\vspace{-4pt}

\newpage

\section{Implications}

%\vspace{-2pt}

%\noindent
This section discusses the implications from this work, including how Mixtures can be connected with others; and the impacts on our understanding on (In)finity.

%\vspace{-4pt}

\subsection{Mixture Meets Other Concepts}

%\vspace{-2pt}

\subsubsection{Mass, Energy and Functionalities}
It is very interesting to observe that: the definition of Mixture partially relates with the changes in terms of the mass for a specific subject. This work conjectures that: the correlations between functionalities and energy may be heavily correlated, and certainly it demands further studies.

\subsubsection{Layout-directed Functionalities, and Orbital Systems} It is also very interesting to observe that: different layouts can impose different functionalities. This also contributes to the understanding of Atomic/Molecular Orbital systems: these stable existence may be useful as reference points\footnote{They can be considered as a part of the pre-established harmony~(\cite{leibniz/monadology}).}.

%\vspace{-4pt}

\subsection{(In)finity, and Leibniz Binary System}

%\vspace{-2pt}

\noindent
Though the author assumes the infinity can be beneficial if there is perpetual motion, this work considers the determination of a finite/infinite can be used for whether a successful formation of Mandelbrot/Julia set can be achieved, under the formalization of~(\cite{ArXiv24/Peng-VRIC}). This is mainly due to its characteristics on the self-similarity, in the context of fractal geometry. If it is finite, it shall be capable to be scaled for the encoding within the Leibniz Binary System~(\cite{leibniz-binary}).

%\vspace{-4pt}

\end{singlespace}

\printbibliography

%%%%%%%%%%%%%%%%%%%%%%%%%%%%%%%%%%%%

\end{document}